\documentclass[submitted,copyright,creativecommons]{eptcs}
\usepackage{breakurl}             
\usepackage{underscore}           
\usepackage{mathptmx}
\usepackage{afterpage}
\usepackage[dvipsnames]{xcolor}
\newcommand\myshade{85}
\colorlet{mylinkcolor}{violet}
\colorlet{mycitecolor}{YellowOrange}
\colorlet{myurlcolor}{Aquamarine}
\hypersetup{ linkcolor  = mylinkcolor!\myshade!black,
             citecolor  = mycitecolor!\myshade!black,
             urlcolor   = myurlcolor!\myshade!black,
             colorlinks = true,
           }
\usepackage{semantic}
\usepackage{verbatim}
\usepackage{enumerate}
\usepackage{listings}
\newcommand{\eqrow}[0]{\stackrel{\scriptsize\textsf{row}}{=\joinrel=}}

\title{An Executable Specification of Typing Rules for\\
  Extensible Records based on Row Polymorphism\\
  {\normalsize\normalfont(Extended Abstract)}}
\author{Ki Yung Ahn
\institute{School of Computer Science and Engineering\\
Nanyang Technological University, Singapore}
\email{yaki@ntu.edu.sg}
}

\begin{document}
\maketitle

Type inference is perceived as a natural application of logic programming (LP).
Natively supported unification in LP can serve as a basic building block of
typical type inference algorithms. In particular, polymorphic type inference
in the Hindley--Milner type system (HM) and its extensions are known to be
succinctly specifiable and executable in Prolog \cite{HMX99,Sulzmann08}.
Our previous work \cite{Ahn2016} demonstrates Prolog specifications of more
advanced features of parametric polymorphism beyond HM such as type-constructor
polymorphism (TCPoly) (a.k.a. higher-kinded polymorphisim). In addition to
typing judgments ($\Delta;\Gamma |- t : \tau$), which relate terms ($t$) to
types ($\tau$), there are kinding judgements ($\Delta |- \tau : \kappa$),
which relate types ($\tau$) to kinds ($\kappa$) in such more advanced
polymorphic type systems. Execution of our Prolog specification for
such a type system (HM+TCpoly) are \emph{multi-staged}, that is,
kinding judgements and typing judgements are executed in separate stages.

In this talk, we demonstrate a specification for typing records, which is one of
the most widely supported compound data structures in real-world programming languages.
More specifically, we develop a specification of a type system that supports
extensible records based on row polymorphism (RowPoly)~\cite{GasJon96}.
We build upon our previous work on HM+TCPoly because the use of rows in record types
are similar to supplying type arguments to a type constructor. For instance,
a type constructor for polymorphic lists ($\textsf{List} : \star -> \star$)
becomes a specific list type (e.g., $\textsf{List Int}: \star$)
when it is supplied a type (e.g., $\textsf{Int}:\star$) as its argument.
Similarly, the record type constructor ($\textsf{Rec} : \textsf{row} -> \star$)
becomes a specific record type
(e.g., $\textsf{Rec}\;\{\textit{name}:\textsf{String},\;\textit{age}:\textsf{Int}\} : \star$)
when it is supplied a row
(e.g., $\{\textit{name}:\textsf{String},\;\textit{age}:\textsf{Int}\}$ : row)
as its argument. 
Similarly to a polymorphic list type $(\forall\alpha:\star.\,\textsf{List}\;\alpha)$
that range over any specific instances of lists, a row-polymorphic record type
$(\forall\rho:\textsf{row}.\,\textsf{Rec}\;\{\textsf{name}:\textsf{String}\mid\rho\})$
ranges over any record type with the \textit{name} field of type \textsf{String};
e.g.,
$\textsf{Rec}\,\{\textit{name}:\textsf{String}\}$
when $\rho$ instantiates to the empty row $\{\}$, or
$\textsf{Rec}\,\{\textit{name}:\textsf{String},\,\textit{age}:\textsf{Int}\}$
when $\rho$ instantiates to $\{\textit{age}:\textsf{Int}\}$.

The extension from HM+TCPoly to HM+TCPoly+RowPoly consists of two aspects.
First is structurally propagating the extension to the kind syntax from
$\,\kappa ::= \star \mid \kappa \to \kappa\,$ to 
$\,\kappa ::= \textsf{row} \mid \star \mid \kappa \to \kappa$.
The syntax of types and terms need to be extended with rows and records, and
the kiding and typing rules for rows and record types need to be added accordingly.
Second is on how to implement those additional rules in Prolog.
Unlike lists, rows are order-irrelevant, for instance,
$\{\textit{name}:\textsf{String},\,\textit{age}:\textsf{Int}\}$ and
$\{\textit{age}:\textsf{Int},\,\textit{name}:\textsf{String}\}$ should be
considered equivalent. Moreover, type inference involving row-polymorphic types
requires unification between rows of unknown size. Unfortunately, Prolog's
native unification supports neither order-irrelevant structures nor
collections of unknown size effectively. We overcome this limitation of
Prolog's unification by implementing a user-defined unification between rows of
unknown size, inspired by the Stolzenburg's work~\cite{Stolzenburg96} on
unification between sets of unknown size.

\begin{table}\centering
\begin{tabular}{l|c|l}
type system       &  size (in 80 col)  &  auxiliary code and extra-logical built-ins used  \\\hline
HM                &  15 lines          &  \verb|\==|, \verb|copy_term|                     \\
HM+TCPoly         &  30 lines          &  DCG related predicates, \verb|gensym|           \\
HM+TCPoly+RowPoly &  45 lines          &  +40 lines of code for row unification, cut (\verb|!|) \\
\end{tabular}
\caption{Size increase, additional auxiliary code and extra-logical built-ins
  used in the Prolog specifications of polymorphic type systems extended from
  HM with TCPoly and RowPoly.}
\label{tbl}
\end{table}

Our type system specifications are self-contained succinct logic programs
with minimal dependencies over external libraries or constraint solving systems,
as summarized in Table~\ref{tbl} below. Each extension of a polymorphic feature adds
about 15 lines of Prolog code for the new typing rules to handle additional
language constructs. In case of RowPoly, we also needed extra 40 lines of
auxiliary Prolog code to implement the unification between rows of possibly
unknown size. This user defined row unification is only required used once in
the specification of the typing rule for applications
$\inference{ \Gamma |- e_1 : A_1 -> B
           & \Gamma |- e_2 : A_2 & A_1 \eqrow A_2}{
             \Gamma |- (e_1 ~ e_2) : B}$
where $\eqrow$ is the type equivalence modulo reordering of row items.
The other typing rules may use the usual structural equivalence.
Further details of our specification for row polymorphic type system
is available on Arxiv \cite{arXiv:1707.07872}.

There are mainly two limitations in our approach of using Prolog to develop
relational executable specification of type systems. First is the use of
extra-logical predicates and meta-programming methods, which makes it difficult
to analyze the specification and argue its correctness. Second is the lack of
error message (except getting ``false'' from Prolog's failure).
For future work, we hope to overcome the first limitation by
(1) formulating a more limited but principled theory of
meta-programming in LP that is suited for specifying multiple levels of
typing rules and (2) by supporting basic operations (e.g., first matching lookup
in a context, row unification) as primitives of the LP system and separately
verify their internal implementation rather than treating as a part of a user
provided specification.
Regarding the second limitation, there are many work on type error messages for
functional languages (e.g., \cite{Stuckey04ited}) but needs further research on
which of the ideas would be compatible with our approach.
\vspace*{-2ex}
\bibliographystyle{eptcs}
\bibliography{main}

\begin{thebibliography}{1}
\providecommand{\bibitemdeclare}[2]{}
\providecommand{\surnamestart}{}
\providecommand{\surnameend}{}
\providecommand{\urlprefix}{Available at }
\providecommand{\url}[1]{\texttt{#1}}
\providecommand{\href}[2]{\texttt{#2}}
\providecommand{\urlalt}[2]{\href{#1}{#2}}
\providecommand{\doi}[1]{doi:\urlalt{http://dx.doi.org/#1}{#1}}
\providecommand{\bibinfo}[2]{#2}

\bibitemdeclare{misc}{arXiv:1707.07872}
\bibitem{arXiv:1707.07872}
\bibinfo{author}{Ki~Yung \surnamestart Ahn\surnameend} (\bibinfo{year}{2017}):
  \emph{\bibinfo{title}{An Executable Specification of Typing Rules for
  Extensible Records based on Row Polymorphism}}.
\newblock \urlprefix\url{https://arxiv.org/abs/1707.07872}.

\bibitemdeclare{inproceedings}{Ahn2016}
\bibitem{Ahn2016}
\bibinfo{author}{Ki~Yung \surnamestart Ahn\surnameend} \&
  \bibinfo{author}{Andrea \surnamestart Vezzosi\surnameend}
  (\bibinfo{year}{2016}): \emph{\bibinfo{title}{Executable Relational
  Specifications of Polymorphic Type Systems Using Prolog}}.
\newblock In: {\sl \bibinfo{booktitle}{{Functional and Logic Programming: 13th
  International Symposium, FLOPS 2016}}}, \bibinfo{series}{{LNCS}},
  \bibinfo{publisher}{Springer International Publishing}, pp.
  \bibinfo{pages}{109–--125}, \doi{10.1007/978-3-319-29604-3_8}.

\bibitemdeclare{techreport}{GasJon96}
\bibitem{GasJon96}
\bibinfo{author}{Benedict~R. \surnamestart Gaster\surnameend} \&
  \bibinfo{author}{Mark~P. \surnamestart Jones\surnameend}
  (\bibinfo{year}{1996}): \emph{\bibinfo{title}{A Polymorphic Type System for
  Extensible Records and Variants}}.
\newblock \bibinfo{type}{Technical Report} \bibinfo{number}{NOTTCS-TR-96-3},
  \bibinfo{institution}{Department of Computer Science, University of
  Nottingham}.

\bibitemdeclare{article}{HMX99}
\bibitem{HMX99}
\bibinfo{author}{Martin \surnamestart Odersky\surnameend},
  \bibinfo{author}{Martin \surnamestart Sulzmann\surnameend} \&
  \bibinfo{author}{Martin \surnamestart Wehr\surnameend}
  (\bibinfo{year}{1999}): \emph{\bibinfo{title}{Type Inference with Constrained
  Types}}.
\newblock {\sl \bibinfo{journal}{Theorey and Practice of Object Systems}}
  \bibinfo{volume}{5}(\bibinfo{number}{1}), pp. \bibinfo{pages}{35--55},
  \doi{10.1002/(SICI)1096-9942(199901/03)5:1<35::AID-TAPO4>3.0.CO;2-4}.

\bibitemdeclare{inproceedings}{Stolzenburg96}
\bibitem{Stolzenburg96}
\bibinfo{author}{Frieder \surnamestart Stolzenburg\surnameend}
  (\bibinfo{year}{1996}): \emph{\bibinfo{title}{Membership-Constraints and
  Complexity in Logic Programming with Sets}}.
\newblock In: {\sl \bibinfo{booktitle}{Frontiers of Combining Systems: First
  International Workshop, Munich, March 1996}}, \bibinfo{publisher}{Springer
  Netherlands}, \bibinfo{address}{Dordrecht}, pp. \bibinfo{pages}{285--302},
  \doi{10.1007/978-94-009-0349-4_15}.

\bibitemdeclare{inproceedings}{Stuckey04ited}
\bibitem{Stuckey04ited}
\bibinfo{author}{Peter~J. \surnamestart Stuckey\surnameend},
  \bibinfo{author}{Martin \surnamestart Sulzmann\surnameend} \&
  \bibinfo{author}{Jeremy \surnamestart Wazny\surnameend}
  (\bibinfo{year}{2004}): \emph{\bibinfo{title}{Improving Type Error
  Diagnosis}}.
\newblock In: {\sl \bibinfo{booktitle}{Proceedings of the 2004 ACM SIGPLAN
  Workshop on Haskell}}, \bibinfo{series}{Haskell '04},
  \bibinfo{publisher}{ACM}, \bibinfo{address}{New York, NY, USA}, pp.
  \bibinfo{pages}{80--91}, \doi{10.1145/1017472.1017486}.

\bibitemdeclare{article}{Sulzmann08}
\bibitem{Sulzmann08}
\bibinfo{author}{Martin \surnamestart Sulzmann\surnameend} \&
  \bibinfo{author}{Peter~J. \surnamestart Stuckey\surnameend}
  (\bibinfo{year}{2008}): \emph{\bibinfo{title}{{HM(X)} Type Inference is
  {CLP(X)} Solving}}.
\newblock {\sl \bibinfo{journal}{Journal of Functional Programming}}
  \bibinfo{volume}{18}(\bibinfo{number}{2}), pp. \bibinfo{pages}{251--283},
  \doi{10.1017/S0956796807006569}.

\end{thebibliography}
\end{document}